\documentclass[aps,pre,reprint,superscriptaddress,nofootinbib]{revtex4-1}
\usepackage{graphicx}
\usepackage{amsmath}
\usepackage{amsfonts}
\usepackage{hyperref}
\usepackage{color}

\renewcommand\vr[1]{\vec{r}_{#1}}
\newcommand{\CC}{C\nolinebreak\hspace{-.05em}\raisebox{.4ex}{\tiny\bf +}\nolinebreak\hspace{-.10em}\raisebox{.4ex}{\tiny\bf +}$\ $}
\newcommand{\CCnospace}{C\nolinebreak\hspace{-.05em}\raisebox{.4ex}{\tiny\bf +}\nolinebreak\hspace{-.10em}\raisebox{.4ex}{\tiny\bf +}}
\newcommand{\cgpu}{\emph{cellGPU} }

\begin{document}

\title{cellGPU: massively parallel simulations of  dynamic vertex models}

\author{Daniel M. Sussman}
\email[]{dmsussma@syr.edu}
\affiliation{Department of Physics, Syracuse University, Syracuse, New York 13244, USA}

\begin{abstract}
Vertex models represent confluent tissue by polygonal or polyhedral tilings of space, with the individual cell interacting via force laws that depend on both the geometry of the cells and the topology of the tessellation. This dependence on the connectivity of the cellular network introduces several complications to performing molecular-dynamics-like simulations of vertex models, and in particular makes parallelizing the simulations difficult. \cgpu addresses this difficulty and lays the foundation for massively parallelized, GPU-based simulations of these models. This article discusses its implementation for a pair of two-dimensional models, and compares the typical performance that can be expected between running \cgpu entirely on the CPU versus its performance when running on a range of commercial and server-grade graphics cards. By implementing the calculation of topological changes and forces on cells in a highly parallelizable fashion, \cgpu enables researchers to simulate time- and length-scales previously inaccessible via existing single-threaded CPU implementations.
\end{abstract}

\maketitle

\section{Introduction}
Simple models that represent confluent tissue by  polygonal or polyhedral tilings of space have a rich history of biophysical application \cite{Brodland2004,Honda2001,Manning2010, Bi2014}. In these ``vertex models'' the degrees of freedom have traditionally been the vertices of the polygons in the tessellation, although some recent models take the degrees of freedom to be the positions of the cells themselves (with the cell geometry following from, e.g., a Delaunay triangulation of those positions) \cite{Bi2016,SAMOS}. Although restricting cell shapes to Voronoi volumes does limit the range of cell configurations relative to more permissive vertex models, this restriction in the spirit of early work proposing Voronoi tessellations as a description of the arrangement of cells in epithelial tissue \cite{Honda1978}.

Regardless of the choice for the underlying degrees of freedom, the models are united by similar equations of motion and force laws that depend on both local geometry and the topology of the cellular network. In this sense vertex models share many similarities with models of soaps and foams \cite{Fletcher2014}, but in the context of living tissues frequently add features like active motion, cell division, and cell death. Vertex models have greatly contributed to the understanding of collective cell growth and migration, which in turn inform processes as diverse as cell sorting \cite{Belmonte2008}, collective cell motion in epithelial tissue \cite{Vicsek2006,Hakim2013}, wound healing \cite{Fredberg2013}, and embryonic development \cite{Hufnagel2006, Farhadifar2007,Staple2010}.

Although the equations of motion used to simulate these models are often quite similar to those used in standard molecular dynamics, a number of computational challenges have prevented the adoption of modern, heavily parallelized implementations common to other simulations. In particular, as noted above, rather than interacting via (typically) short-ranged forces vertex models have interactions that depend on both local geometry \emph{and} the topology of the interacting units. For example, models that represent cells as Voronoi volume elements \cite{Bi2016,SAMOS} clearly require that a either a Voronoi or Delaunay triangulation be maintained at all times. From that triangulation the cell shapes and cell neighbors are defined, which in turn are the direct inputs to determining the forces which enter the equations of motion governing the model.

Two important consequences of this stand out. First, although the computational cost of both (1) performing a Delaunay triangulation on a point set of size $N$ and (2) advancing a standard molecular dynamics algorithm by one time step given $N$ particles interacting via short ranged forces scale like $\mathcal{O}\left(N\log N\right)$, the prefactor for the Delaunay triangulation is typically \emph{much} worse than the prefactor for simulating, e.g., Lennard-Jones particles with a reasonable finite-range cut off \cite{CGALTriangulation, Plimpton1995}. Second, and more seriously in light of modern trends in scientific computing, the triangulation problem is much less amenable to parallelized computation. Indeed, a recent review has explicitly highlighted the lack of parallelization in these cell models as the main bottleneck to further research \cite{Drasdo2015}, and even the most sophisticated open-source implementations of vertex models currently rely on serial, single-threaded CPU-based implementations \cite{CHASTE, SAMOS}. Because of this, vertex models struggle to tackle both long-timescale problems and simulations of large numbers of interacting cells. The latter is particularly problematic for two-dimensional vertex models, in which finite-size effects may be especially strong.

Written in \CCnospace/CUDA \cite{CUDA}, \cgpu lays the foundation for moving to highly parallelized, GPU-based simulations of these models. Two classes of vertex models are described in this article: a fully GPU-accelerated ``active vertex model'' (AVM) in which vertices move according to both forces and the intrinsic activity of nearby cells, and a hybrid GPU/CPU implementation of a Delaunay-triangulation-based model of active cells (the ``self-propelled Voronoi'' (SPV) model \cite{Bi2016}). The remainder of the paper is organized as follows. Sec. \ref{sec:imp} begins with a more in-depth overview of the two vertex models mentioned above, and then describes the details of the parallelized implementation. Section \ref{sec:results} provides performance benchmarks, comparing the GPU-accelerated algorithms with existing CPU-based implementations. Section \ref{sec:disc} closes with a discussion of future directions for the code, and has additional information on its availability.

\section{Design and implementation\label{sec:imp}}

\subsection{Vertex model overview}
\begin{figure}
\begin{center}
\includegraphics[width=.32\linewidth]{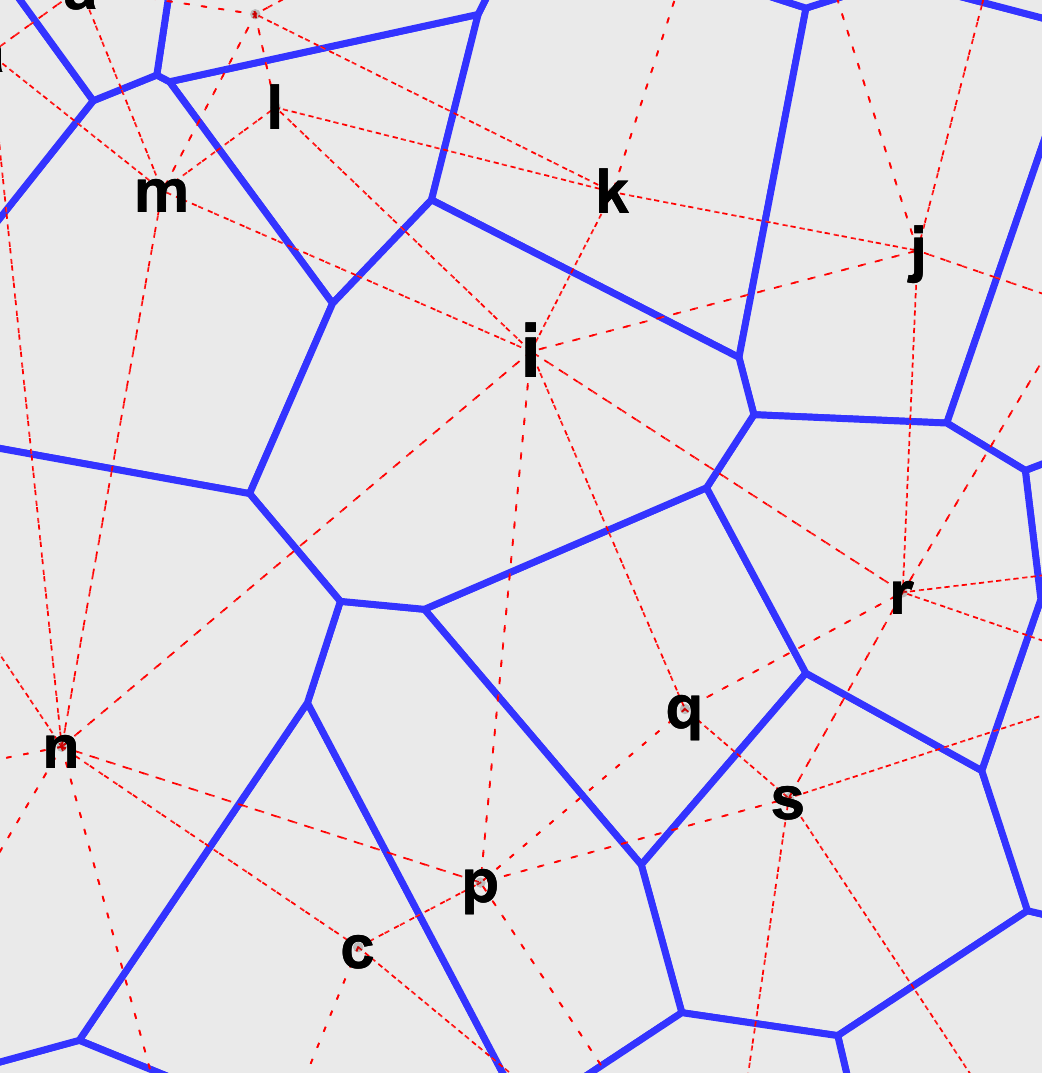}\hfill
\includegraphics[width=.6\linewidth]{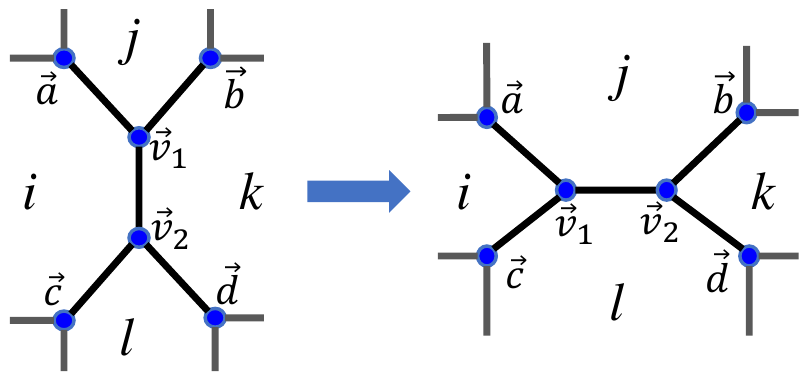}
\caption{(Left) A self-propelled Voronoi model, in which each cell is a Voronoi area, shown in blue solid lines. Interactions depend on both the cell shapes and the connectivity of the dual Delaunay triangulation of the cell positions, which is shown in dotted red lines. (Right) A T1 transition, in which the edge connecting vertices $\vec{v}_1$ and $\vec{v}_2$ is rotated and an exchange of cell neighbors occurs, with cells $i$ and $k$ adjacent before the transition and cells $j$ and $l$ adjacent after the transition. The active vertex model has the vertices as the degrees of freedom and cellular topology evolves via these (and other) transitions.}
\label{fig:cartoon}
\end{center}
\end{figure}

In two dimensions vertex models describe a confluent patch of tissue as a polygonal tiling of space, each polygon corresponding to a coarse-grained representation of a cell. The degrees of freedom are either the collection of vertex positions, here denoted as $\vec{h}_i$, or the positions of the cells themselves, denoted as $\vec{r}_i$. In the SPV model each cell is represented not by a generic polygonal region, but by a Voronoi area element, and so the topology of the network refers to an underlying Delaunay triangulation of the cell positions. In contrast, in the AVM the vertices themselves are the degrees of freedom, and changes to the topology occur via cell neighbor exchange (T1 transitions), cell removal or addition (T2 transitions), vertex-edge merging (T3 transitions) \cite{Fletcher2014}. Schematic illustrations of these models are shown in Fig. \ref{fig:cartoon}.

The forces acting on the degrees of freedom in these simplified cell models are derived by first defining a simple energy functional for the entire tissue, for example
\begin{equation}\label{eq:energy}
E = \sum_{i=1}^N K_A(A_i - A_{0,i})^2 + K_P(P_i-P_{0,i})^2.
\end{equation}
Here $N$ is the total number of cells, $A_i$ and $P_i$  are the area and perimeter of cell $i$, $A_{0,i}$ and $P_{0,i}$ are ``preferred'' values of the area and perimeter for cell $i$, and $K_A$ and $K_P$ are area and perimeter moduli. Many variations of the above functional have been written; to be concrete we discuss the implementation of the simple one above. Here the quadratic dependence on cell area can be interpreted as a cell monolayer's resistance to height fluctuations due to adhesions between cells and cell incompressibility, and the quadratic dependence on cell perimeter as a competition between active contractility of the actomyosin sub-cellular cortex and tension due to both cell-cell adhesions and cortical tension \cite{Bi2016}. From the energy, the force on a degree of freedom (either a vertex or a cell position), $j$, is defined by 
\begin{equation}\label{eq:force}
\vec{F}_j = \nabla_j E.
\end{equation}
Note that energy functional in Eq.~\ref{eq:energy} leads to forces which are fundamentally many-body (not pairwise) in nature.

In addition to these intercellular forces, active versions of these models include additional cell motility terms. For instance, in the spirit of simple models of self-propelled particles \cite{Henkes2011}, the SPV model assigns a polarization vector to every cell, $\vec{n}_i = (\cos \theta_i, \sin \theta_i)$, and posits that cells exert a self-propulsion force of constant magnitude $v_0/mu$, where $\mu$ is the mobility (with units of an inverse frictional drag) and $v_0$ is a self-propulsion speed \cite{Bi2016}. The polarization vector of every cell rotates randomly according to $\partial_t\theta_i = \eta_i(t)$, where $\eta_i$ is a white-noise process with zero mean and variance $2D_r$; the rotational diffusion constant $D_r$ in the model determines a time scale of persistent motion that would arise in the absence of other interactions. Taken together, then, the SPV equation of motion for cell $i$ is \cite{Bi2016}
\begin{equation}\label{eq:spv}
\frac{d\vec{r}_i}{dt} = \mu \vec{F}_i + v_0\vec{n}_i.
\end{equation}
We define the active vertex model by implementing the self-propulsion at the level of cells and taking the equation of motion for the vertices to be 
\begin{equation}\label{eq:avm}
\frac{d\vec{h}_i}{dt} = \mu\vec{F}_i + \frac{v_0}{n}\sum_{\langle i j\rangle}\vec{n}_j,
\end{equation}
where $\langle ij \rangle$ represents the $n$ cells indexed by $j$ that are neighbors of vertex $i$. That is, each cell has a constant self-propulsion as in the SPV, and the activity of each vertex is a simple average of the activity of the adjoining cells.

\subsection{Computational overview}
We first give an overview of the general computational process that is common to both models described above, breaking down the simulation time step into a sequence of logically distinct calculations. Since the forces in the models depend on the geometry of the cells at the beginning of the time step, the area and perimeter of every cell must first be computed. Following that, the forces on the degrees of freedom must be evaluated from Eq.~\ref{eq:force}, and the appendix shows in detail how to calculate these forces in a computationally efficient manner. Since computing the many-body forces from Eq.~\ref{eq:force} is expensive relative to, e.g., evaluating pairwise Lennard-Jones interactions, even parallelizing these geometry and force calculations can lead to substantial performance gains.

With the forces in hand the degrees of freedom can be displaced according to either Eq.~\ref{eq:spv} or \ref{eq:avm}. The next challenge is determining whether the cellular topology needs to be updated; implementing vertex models in a highly parallelizable fashion has required abandoning some of the data structures commonly used to represent the topology of cellular networks (or, more generally, to represent collections of vertices, edges, and cells). For example, the doubly connected edge list (DCEL) is a particularly convenient data structure when working on the CPU \cite{Muller1978}. In that representation every edge is composed of two ``half-edges,'' each of which has pointers to the face that it bounds, the vertex that serves as its origin, and the other half-edge making up the edge (the half-edge's ``twin''). This sort of structure makes it easy to update a triangulation or traverse it in some desired order, but at the cost of having very non-local memory accesses. Even on the CPU this can lead to significant performance costs, where if not carefully implemented the resulting cache-access patterns can change the effective observed algorithmic complexity of vertex model simulations. On the GPU these non-local memory requests can be even more of an imposition, and the performance of accessing and operating on data is radically enhanced by working with much flatter data structures in which the majority of memory requests are local. 

Thus, \cgpu takes the approach of initializing and then maintaining many redundant data structures representing the topology -- such as a one-dimensional list keeping track of vertices around each cell in counter-clockwise order, a separate list keeping track of the vertices each vertex is connected to, and yet a third list keeping track of the cells that border each vertex. The code has been profiled to confirm that the cost of maintaining these redundant data structures is much less than the performance gains that come with their use. These performance considerations will be revisited in Sec. \ref{sec:results}.

Given its relatively simple rules for when topological transitions can occur, the AVM branch of \cgpu can manage the step of checking and updating the topology entirely on the GPU, avoiding the need for costly memory transfers between the host (CPU) and device (GPU). The SPV branch of \cgpu instead adopts a hybrid approach of testing the topology fully on the GPU but performing any necessary repairs on the CPU. Eventually, though, even this step will be moved to the GPU, and algorithmic choices have been made (discussed below) to facilitate this eventual move. Note that the current implementation of all CPU routines in \cgpu is strictly single-threaded, although here too algorithmic choices have been made so that multithreaded re-implementations would be straightforward.

As a final detail, note that in these simulations there is a strong incentive to try to order all of the data arrays so that particles which are physically close to each other are also close to each other in memory. This is a consequence of the slow memory access on the GPU, and the improvements to memory access speeds that can come if all read/write operations are coalesced \cite{CUDA}. That is, a thread that has to read data from many different parts of a given array -- e.g., reading off the vertex positions for cells close to a given cell from an un-ordered array of vertex data in order to test the emptiness of a circumcircle -- would be much slower than a thread which could read the same information from a contiguous block of the array. 

As is common in modern particle-based simulations \cite{Plimpton1995,Anderson2008}, \cgpu uses a Hilbert-curve-based spatial sorting scheme. This takes an approximation to the Hilbert curve (a particular space-filling fractal curve chosen for its optimal data-locality properties \cite{Moon2001}), finds the location on the curve closest to each cell, and sorts cells based on the order in which they occur along the approximation to the Hilbert curve. As demonstrated in Sec. \ref{sec:results} for the AVM, this represents a (small) optimization even for the CPU branch of the code, but radically improves performance on the GPU. Thus, \cgpu provides the option to periodically resort the degrees of freedom, which should be done at a frequency related to how much the cells move and rearrange neighbors.

\subsection{Self-propelled Voronoi model implementation}
As noted above, \cgpu adopts a hybrid GPU/CPU approach to simulating the two-dimensional SPV model defined by Eq.~\ref{eq:spv}. The hybrid approach is currently required to maintain the Delaunay triangulation, but future development is planned to eventually implement the model simulation entirely on the GPU. A highly parallelizable approach to testing the validity of the triangulation has already been implemented, and algorithms for maintaining the triangulation have been chosen with an eye towards their eventual implementation on the GPU. The following sub-sections highlight the sequence of GPU and CPU function calls that form the basic loop of advancing the simulation by one time step.

\subsubsection{Computing the geometry of cells}
At the beginning of a simulation time step, the geometry of every cell (i.e. the current area and perimeter of each cell) is calculated on the GPU as a prerequisite to computing the forces on the cells. The parallelism here is at the level of independently computing the geometry of each cell, and a different GPU thread can be assigned to each cell index. In the course of computing the geometry, the kernel computes the Voronoi vertices that form the boundary of each cell by evaluating the circumcenter of every triangle in the Delaunay triangulation (we reiterate that in the AVM the vertices themselves are the degrees of freedom -- they do not need to form a Voronoi configuration, and computing the geometry of the cells must be done very differently). These arrays of Voronoi positions are saved for later use (after determining that the cost of saving and loading these Voronoi vertices is less than the cost of recomputing them on the fly).

\subsubsection{Calculating the forces on cells}
With the geometry computed and the Voronoi vertex positions saved, the force on every cell is computed. Note that additional parallelism can be exposed here: as shown in the appendix, the net force on a cell can be decomposed into the contribution to that force from each of the cell's Voronoi vertices. Thus, the force calculation step is broken into two GPU kernel calls. First ``force sets'' are computed, so that a different thread computes the force on cell $i$ due to vertex $v$ for each $i$ and $v$ independently. Second, a different kernel takes this data, together with neighbor lists containing information on which vertices are associated with which cell, and assigns one thread per cell to add the individual force set contributions to get the net force on each cell.

\subsubsection{Displacing particles and testing the topology}
A straightforward kernel takes the net force on each cell, updates the cell positions according to both these net forces and the self-propulsion of each cell, and rotates the cell directors via a GPU-based pseudo-random number generator.

Next, the triangulation must be maintained. Depending on the parameter values of the SPV model that are chosen, it is entirely possible that the network topology -- the connection between cells and vertices -- remains unchanged. One approach to solving this problem is based on constructing the digital Voronoi diagram to test for changes in the topology of the triangulation \cite{Bernaschi2016}. Here, instead, we directly test the empty-circumcircle property that defines the Delaunay triangulation. Each cell in the SPV model corresponds to a Delaunay vertex, and by definition a valid Delaunay triangle has the property that a circumcircle defined by the vertices of that triangle contains no other vertices. Thus, a kernel is launched in which each thread checks the circumcircle defined by one of the triangles in the Delaunay triangulation to see if any cell has moved in such a way that an update of the triangulation is needed.

\subsubsection{Repair Delaunay triangulation as needed}
The kernel that tests the empty-circumcircle property copies a flag back to the host CPU, indicating whether any of the circumcircle tests failed. If no tests failed (i.e. the topology of the Delaunay triangulation did not change as the cells moved during that time step) then the next time step can be immediately started on the GPU. If a circumcircle came back non-empty, though, a list of cells that need an update to their neighbor list is returned to the CPU. For each of these flagged cells, the triangulation is \emph{locally} repaired. Typically this would be done via an edge-flipping or star-splaying algorithm \cite{Shewchuk2005}, and such algorithms form the basis for existing hybrid CPU/GPU triangulation routines for computing large-scale Delaunay meshes \cite{Cao2014}.

As an alternative, and with the aim of eventually implementing the triangulation repair step on the GPU and eliminating the need for a hybrid algorithm, a different approach is followed. We first note a locality lemma that can be used to limit the possible set of Delaunay neighbors of a given cell: given a set of Delaunay vertices that form a polygon bounding the cell $i$, the set of Delaunay neighbors of $i$ is a strict subset of the cells that are contained in the circumcircles defined by $i$ and any two consecutive vertices of on the bounding polygon \cite{Chen2012}. These points form the candidate ``Delaunay 1-ring'' of cell $i$. This set is easily found on the CPU, and the data is handed off to an efficient CGAL library to reduce the candidate 1-ring to the actual set of neighbors of the cell. One advantage of this algorithm is that even on the CPU it can be readily parallelized to find the candidate 1-ring for all cells that need updating simultaneously. When all flagged cells have been repaired, the auxiliary data structures describing the topology of the triangulation are updated, and all data is sent back to the GPU.

\subsection{Active vertex model implementation}
Implementing the AVM described by Eq.~\ref{eq:avm} is, in some ways, more straightforward than the hybrid implementation of the SPV model. Many of the components of the simulation time step are carried out exactly as described above, beginning with the calculation of the geometry of every cell and moving on to the net force on each vertex. The kernel for displacing vertices is modestly more complicated, since each vertex needs to know about the director defining the self-propulsion of the vertex' neighboring cells. This is optimized by having separate kernel calls to first move the vertices and then update the cell directors.

The advantage of simulating the AVM model is that maintaining the topology is much simpler. As mentioned above, rather than referencing an underlying Delaunay triangulation, topological changes in the AVM are managed by T1, T2, and T3 transitions. At present \cgpu only implements T1 transitions, but all three transition types are readily compatible with being performed only via local access to the auxiliary data structures used to define the topology. As such, the topological changes can be carried out entirely on the GPU, removing the need for hybrid CPU/GPU operation. The primary difficulty is the need to avoid race conditions that could be caused by simultaneously executing multiple topological transitions involving the same vertex or cell. To deal with this, \cgpu adopts a strategy of only allowing a single topology update per cell per kernel call. All of the topological changes called for in a given time step are guaranteed by calling the topology-updating kernel repeatedly, i.e., until no more changes are called for.

\section{Performance benchmarks\label{sec:results}}

\begin{figure}
\begin{center}
\includegraphics[width=1\linewidth]{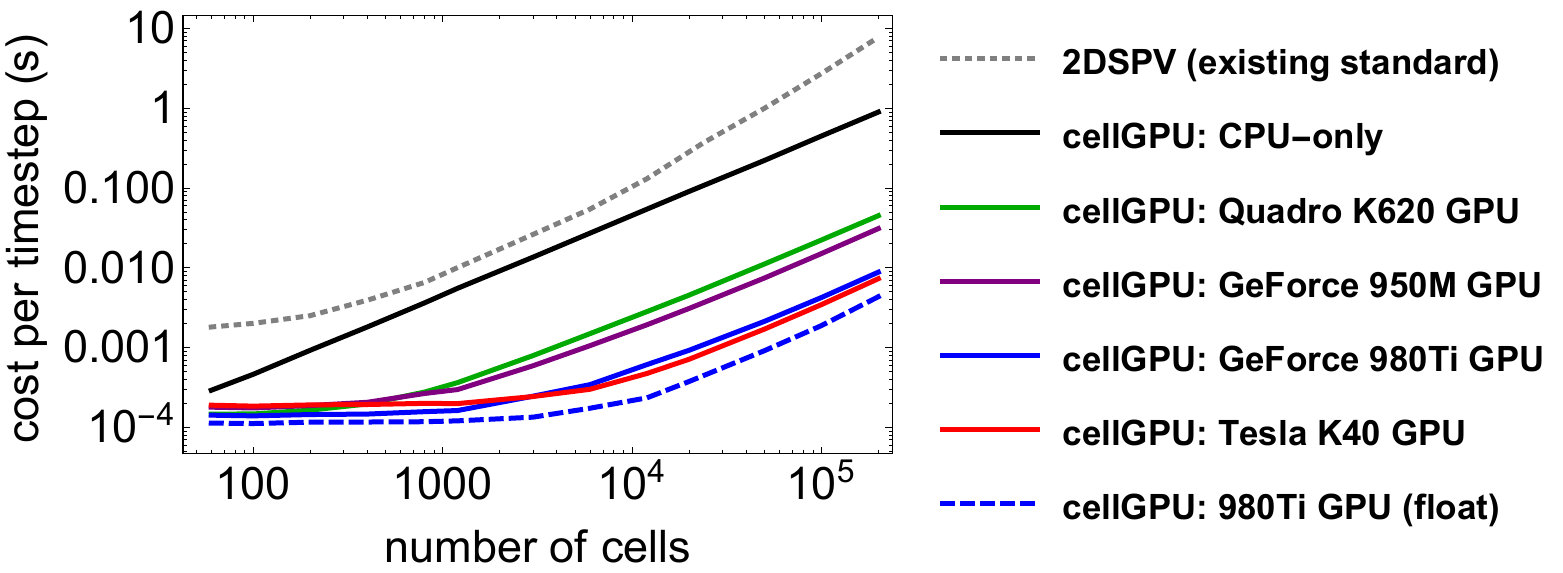}
\caption{Computational performance of \cgpu for SPV model parameters in the solid-like regime ($P_0=3.8$, $v_0=0.01$) with a simulation time step of $\Delta t = 0.05$. From top to bottom, the gray curve corresponds to a standard CPU implementation of the SPV, based on doubly-connected edge lists (DCEL), and the black curve corresponds to \cgpu running only on a single CPU thread. The remaining curves compare the algorithm's performance when running on different graphics cards: in order the green curve corresponds to using on a Quadro K620, the purple to a GeForce 950M, the blue solid curve to a GeForce 980Ti, and the red to a Tesla K40. All of these curves represent calculations in double precision; the dashed blue curve shows the performance on a GeForce 980Ti when floating-point precision is used instead. }
\label{fig:spvCardTiming}
\end{center}
\end{figure}

Benchmarking of \cgpu was carried out on two systems: (1) a workstation running Ubuntu 14.04 with a 3.5 GHz  Xeon E5-1620 V3 processor with 32 GB of RAM with the option of using either a Tesla K40, Quadro K620, or GeForce 980Ti graphics card, or (2) a laptop running Ubuntu 16.04 with a 2.2 GHz Core i5 5200U processor with 8 GB RAM and a GeForce 950M graphics card. Figure \ref{fig:spvCardTiming} shows the performance of \cgpu when simulating the SPV model in a favorable regime, i.e., in the solid-like regime where cells move slowly and the Delaunay triangulation only needs to be updated occasionally. The performance is shown for a range of consumer- and server-grade graphics cards; as can be seen, the performance can be as much as three orders of magnitude faster than existing CPU-based implementations for large system sizes. The super-linear scaling seen in the DCEL-based implementation results not from true a difference in algorithmic complexity but from the effect of repeated cache misses when accessing data scattered throughout memory.

\begin{figure}
\includegraphics[width=1\linewidth]{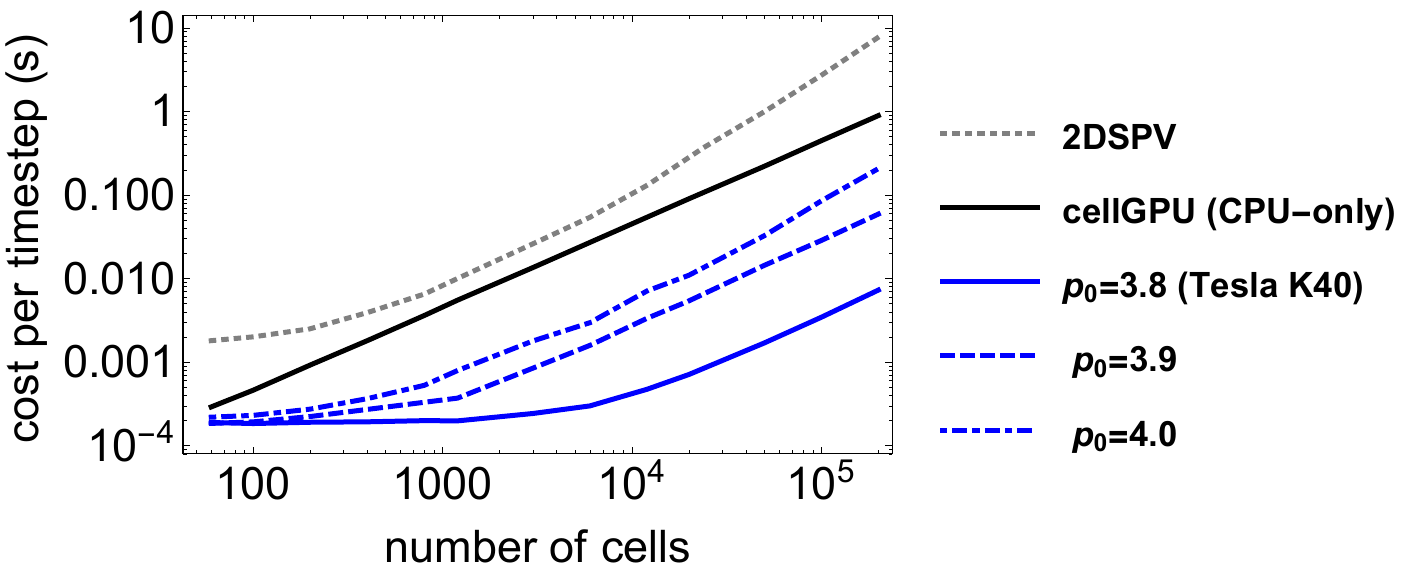}
\caption{Performance of \cgpu for the SPV model at fixed $v_0$ and but varying $P_0$ from the solid to fluid regimes. The top two curves are again the CPU-based implementations, which on this scale have a computational speed that is independent of simulation parameters. From bottom to top the solid blue curve corresponds to $P_0=3.8$, the dashed blue curve to $P_0=3.9$, and the dot-dashed blue curve to $P_0=4.0$, all running on a Tesla K40 graphics card.}
\label{fig:spvP0timing}
\end{figure}

When topological changes happen more frequently, the hybrid implementation results in less dramatically accelerated performance. As the model is simulated in the increasingly fluid-like regime cells move much more and the topological updating scheme needs to be invoked on the CPU more and more frequently. In addition to the cost of locally repairing the triangulation, this also causes expensive data transfers between the host and the device. This overall degradation of performance can be seen for a particular choice of parameters in Fig. \ref{fig:spvP0timing}.

\begin{figure}
\begin{center}
\includegraphics[width=1\linewidth]{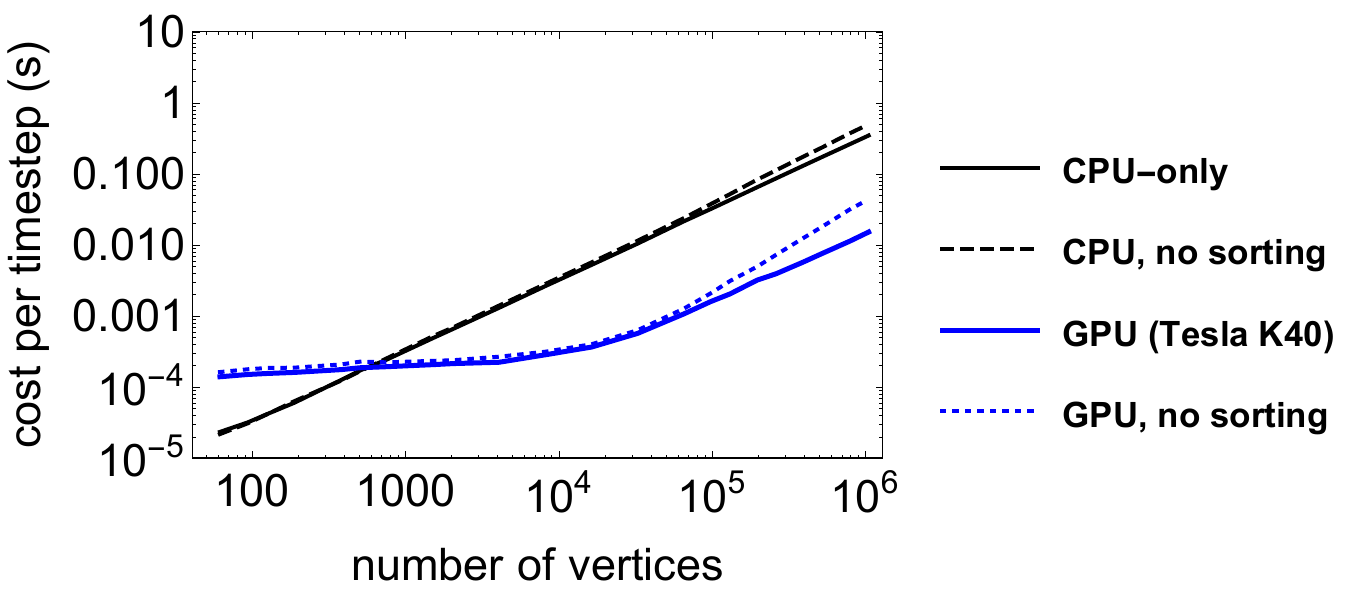}
\caption{Performance of \cgpu AVM implementation. The linear black curves correspond to the strictly CPU implementation, and the blue curves correspond to the GPU-accelerated implementation running on a Tesla K40. For each set of curves the dashed curve shows the performance when no Hilbert sorting scheme is used.}
\label{fig:avmtiming}
\end{center}
\end{figure}

In contrast, since the AVM branch of the \cgpu code is fully GPU accelerated it has almost no dependence on the parameter regimes in which the simulations are performed. Figure \ref{fig:avmtiming} shows the performance of the simple active vertex model implemented by \cgpu, which can again be more than an order of magnitude faster than its own CPU-based implementation, and multiple orders of magnitude faster than non-\CC implementations still in common use \cite{Manning2010}. The figure also shows the importance of using the spatial sorting schemes described above to maintain data locality. While Hilbert-curve sorting has only a modest impact on the CPU branch of the code, neglecting it on the GPU branch causes a change in the effective observed computational complexity of the model, from roughly linear in the number of vertices simulated per time step to a distinctly super-linear scaling.

\section{Conclusions \label{sec:disc}}
We have demonstrated the computational gains that come from implementing molecular dynamics simulations of off-lattice cell models in a highly parallelized manner, allowing access to simulation time- and length scales unavailable to existing single-threaded implementations. Natural extensions include incorporating more general sets of boundary conditions and supporting additional classes of off-lattice cell models, including extensions to three-dimensional models. Additionally, work to move the self-propelled Voronoi branch away from its current hybrid implementation and towards a fully GPU-accelerated algorithm is currently planned. Code availability and more details about the planned future directions of the code and developments that are currently underway can be found at \href{https://gitlab.com/dmsussman/cellGPU}{Ref. \cite{cellGPU}}, with additional documentation maintained at \href{http://dmsussman.gitlab.io/cellGPUdocumentation/}{http://dmsussman.gitlab.io/cellGPUdocumentation}.

\begin{acknowledgments}
I would like to thank Lisa Manning, Matthias Merkel, Michael Czajkowski, and David Yllanes for fruitful discussions and comments on this manuscript. This work was supported by NSF-POLS-1607416; the Tesla K40 used for this research was donated by the NVIDIA Corporation.
\end{acknowledgments}

\appendix
\section{Efficiently computing vertex model forces}
\begin{figure}[htbp]
\begin{center}
\includegraphics[width=0.45\linewidth]{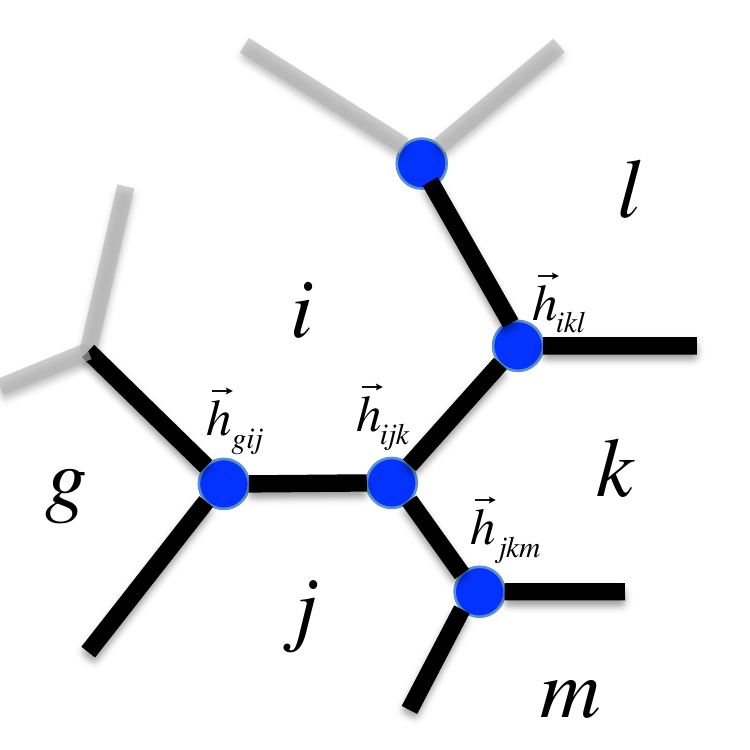}
\caption{Schematic diagram of cell $i$ and some of its neighbors, along with the associated Voronoi vertices (the circumcenters of three adjacent cells) labeled for convenient reference.}
\label{fig:forceSchematic}
\end{center}
\end{figure}

If the topology of the cellular network is known -- either from an underlying triangulation of space or via a direct enumeration of vertex-vertex connections -- computing the forces on the degrees of freedom of a vertex model is straightforward. For reference, Fig. \ref{fig:forceSchematic} provides a schematic picture of a relevant patch of a two-dimensional tissue model. To be explicit, each vertex is labeled by the three cells it is adjacent to.

\subsection{AVM forces}
Recall that the energy functional has the form
\begin{equation}
E = \sum_i^N\left( K_A (A_i-A_0)^2 + K_P(P_i-P_0)^2\right),
\end{equation}
and that the force on vertex $\vec{h}_{ijk}$ is
\begin{equation}
-\frac{\partial E}{\partial \vec{h}_{ijk} } =-\left( \frac{\partial E_i}{\partial \vec{h}_{ijk} } + \frac{\partial E_j}{\partial \vec{h}_{ijk} } + \frac{\partial E_k}{\partial \vec{h}_{ijk} } \right),
\end{equation}
i.e., the motion of vertex $\vec{h}_{ijk}$ only changes the shape of cells $i,$ $j,$ and $k$. The energy derivatives are straightforward, e.g., 
\begin{equation}
\frac{\partial E_i}{\partial  \vec{h}_{ijk}} = 2 K_a(A_i-A_0)\frac{\partial A_i }{\partial  \vec{h}_{ijk}}+2 K_P(P_i-P_0)\frac{\partial P_i }{\partial  \vec{h}_{ijk}} ,
\end{equation}
and where the area and perimeter derivatives with respect to Voronoi vertex positions can themselves be written as follows. Let $\vec{t}_{ij} = \vec{h}_{ijk} -  \vec{h}_{gij}$, $\vec{t}_{ik} =   \vec{h}_{ikl}- \vec{h}_{ijk} $, with $\hat{t}_{ij}$ and $\hat{t}_{ik}$ being the unit vectors in those directions. Similarly, let $l_{ij}$ be the length of the edge between cell $i$ and $j$, and $\hat{n}_{ij}$ be the unit vector pointing outwardly normal to that cell edge. Then one can write
\begin{eqnarray}
\frac{\partial A_i }{\partial \vec{h}_{ijk}} &= &\frac{1}{2}\left( l_{ij} \hat{n}_{ij} +l_{ik} \hat{n}_{ik}   \right) \\
\frac{\partial P_i }{\partial \vec{h}_{ijk}} &= &-\left(\hat{t}_{ij}+\hat{t}_{ik}  \right)
\end{eqnarray}

\subsection{SPV forces}
Following Bi et al. \cite{Bi2016}, the force on cell $i$ in cartesian direction $\lambda$ can be computed as 
\begin{equation}
F_{i\lambda}=-\frac{\partial E}{\partial r_{i\lambda}} = -\frac{\partial E_i}{\partial r_{i\lambda}}-\sum_{<ij>}\frac{\partial E_j}{\partial r_{i\lambda}},
\end{equation}
where $\langle ij \rangle$ refers to all cell neighbors $j$ of cell $i$, referring to the configuration in Fig. \ref{fig:forceSchematic}. The terms in the above can be expanded via the chain rule, for instance:
\begin{equation}
\frac{\partial E_k}{\partial r_{i\lambda}} = \sum_{\nu}\left(\frac{\partial E_k}{\partial h_{ijk,\nu}}\frac{\partial h_{ijk,\nu}}{\partial r_{i\lambda}} +\frac{\partial E_k}{\partial h_{ikl,\nu}}\frac{\partial h_{ikl,\nu}}{\partial r_{i\lambda}} \right).
\end{equation}
Here these are the only terms needed, since the other voronoi vertices associated with cell $k$ (the middle of the three neighboring cells in clockwise order) do not depend on the position of cell $i$. The partial derivatives here depend on the positions of $\vec{h}_{jkm}$ and $\vec{h}_{kln}$, where $m$ is the cell other than $i$ that has both $j$ and $k$ as neighbors, and $n$ is the cell other than $i$ that has both $k$ and $l$ as neighbors, so that the forces in the SPV model depend on nearest and next-nearest Delaunay neighbors of cell $i$. The derivative of the energy with respect to the voronoi vertices was calculated above.

The derivatives of the Voronoi vertices with respect to the position of the cell, e.g. $(\partial \vec{h}_{ijk})/(\partial \vr{i})$, can be calculated efficiently as follows. Let $\vr{ij}$ denote the vector from $i$ to $j$, and define:
\begin{eqnarray}
c & = & \vec{r}_{ij,x}\vec{r}_{kj,y} - \vec{r}_{ij,y}\vec{r}_{kj,x}\\
d & = & 2 c^2\\
\vec{z} & = & \beta d \vr{ij} + \gamma d \vr{ik}\\
\beta d & = & -\left|\vr{ik}\right|^2 \cdot \left(\vr{ij}\cdot\vr{jk}\right)\\
\gamma d & = & \left|\vr{ij}\right|^2 \cdot \left(\vr{ik}\cdot\vr{jk}\right).
\end{eqnarray}
These expressions are related to how the positions of the Voronoi vertex was written in Ref. \cite{Bi2016}. Further writing the derivatives 
\begin{eqnarray}
\frac{\partial (\beta d)}{\partial \vr{i}} & = & 2 \left(\vr{ij}\cdot\vr{jk} \right)\vr{ik} +\left|\vr{ik}\right|^2 \vr{jk} \\
\frac{\partial (\gamma d)}{\partial \vr{i}} & = & -2 \left(\vr{ik}\cdot\vr{jk} \right)\vr{ij} +\left|\vr{ij}\right|^2 \vr{jk} \\
\frac{1}{d}\frac{\partial d}{\partial \vr{i}} & = & \frac{2}{c} \left\{-\vec{r}_{jk,y},\vec{r}_{jk,x}  \right\}.
\end{eqnarray}
Finally, with $I_2$ representing the $2\times 2$ identity matrix and $\otimes$ the dyadic product, the desired change in Voronoi vertex position with respect to cell position is 
\begin{widetext}
\begin{equation}
\frac{\partial \vec{h}_{ijk}}{\partial \vr{i}} =I_2+ \frac{1}{d}\left[ \vr{ij}\otimes \left(\frac{\partial (\beta d)}{\partial \vr{i}}\right) + \vr{ik}\otimes \left(\frac{\partial (\gamma d)}{\partial \vr{i}}\right) - (\beta d+\gamma d)I_2 - \vec{z}\otimes \left( \frac{1}{d}\frac{\partial d}{\partial \vr{i}}\right)   \right].
\end{equation}
\end{widetext}

\bibliography{cellGPU_bib}

\end{document}